# USING THE CORRELATION BETWEEN ISGRI SURVEY RESULTS AND OTHER CATALOGUES TO AID SOURCE IDENTIFICATION

**John B. Stephen, A. Malizia and L. Bassani**

*IASF-INAF, Via P. Gobetti 101, 40129 Bologna, ITALY, E-mail: stephen@iasfbo.inaf.it*

**ABSTRACT**

In order to help in the identification of INTEGRAL/ISGRI sources, we have developed a software package which allows a rapid and detailed cross-correlation to be performed between various source catalogues. It allows subsets of catalogues to be constructed at a selected level of probability of association, which can then be used to construct multi-waveband correlations.

## 1. INTRODUCTION

It has already been shown [1,2] that cross-correlation analysis of the ISGRI 1st [3] and 2nd [4] surveys with the ROSAT catalogues can be very useful in identifying possible X-ray counterparts which can then, by means of their generally much smaller error boxes, be utilised to pinpoint any optical association. Furthermore, the form of the correlation function, which should follow the point source location accuracy (PSLA) of the IBIS/ISGRI instrument (convolved with the PSLA of the secondary instrument) for true associations, can be used in conjunction with the number distributions to investigate the probability of correlation. In order to extend this work to other wavelengths we have created a tool which allows simple access to all available HEASARC [5] (using the .tdat format) and other catalogues in order to facilitate analysis of the cross-correlations. We are currently using this tool to help identify specific classes of objects.

## 2. THE USER INTERFACE

The graphical user interface (GUI) is written using the Interactive Data Language (IDL) [6], and whilst optimized for use under Windows XP it will also run under Linux. There is a unique GUI which allows the user to :
- Load two data files. The source distributions are shown in the two panels on the left.
- Select subsets of the data files by relative strength (i.e. from a% to b% of the intrinsic source strength range) if this data is available in the catalogue.
- Calculate and display the cross-correlation function.
- Display the positions of those sources in the first data file which either have, or do not have an association in the second date file.
- Calculate and display the theoretical cross-correlation function for the source numbers under various (selectable) distribution assumptions.
- Calculate and display a 'chance' cross-correlation function obtained by mirroring the sources in the first catalogue in a combination of right ascension, declination, galactic longitude and latitude.
- Fit a combination of inverse Gaussian functions to the correlation function in order to investigate the compatibility with the IBIS/Other Catalogue PSLA.
- Toggle between the probability and number of associations as a function of distance
- Output two catalogues containing the subsets of sources in the first catalogue with and without associations in the second catalogue at a user-selected probability or distance.

A typical example is shown in figure 1.

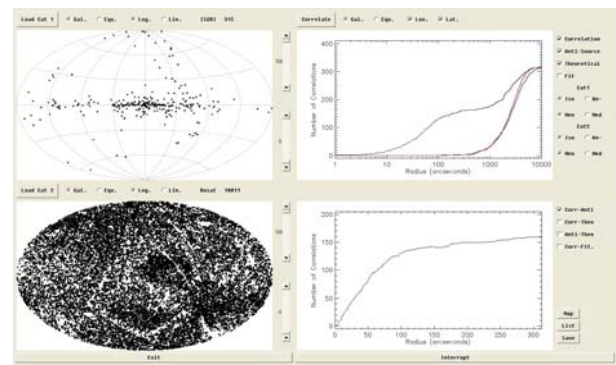

Fig. 1 The correlation analysis of the second IBIS survey catalogue with the ROSAT Bright Source Catalogue.

## 3. THE "2+" IBIS/ISGRI CATALOGUE

The IBIS/ISGRI catalogue which we are currently using consists of the 2nd survey catalogue combined with the Bassani AGN catalogue [7] plus all other sources detected above 5σ significance and reported in the literature before March 2006, for a total of 315 objects. We show below some preliminary results from this analysis. As soon as it becomes available, the 3rd survey catalogue will be employed.

## 4. OTHER CATALOGUES

We show below some preliminary results from this analysis using several catalogues covering a wide range of energies.

### 4.1 X-RAY: The ROSAT Catalogues

The primary catalogues used are those of the ROSAT Bright [8] and Faint [9] Source Catalogues, containing over 18000 and 100,000 sources respectively. As expected there is a clear correlation in both with around 140 sources (~45%) having counterparts in the former and 38 in the latter. Figs 1 and 2 show the graphical user interface (GUI) during the correlation with both these catalogues respectively. Follow up work on the same correlations performed with the 1st and 2nd catalogues has already led to the optical identification of a significant number of sources.

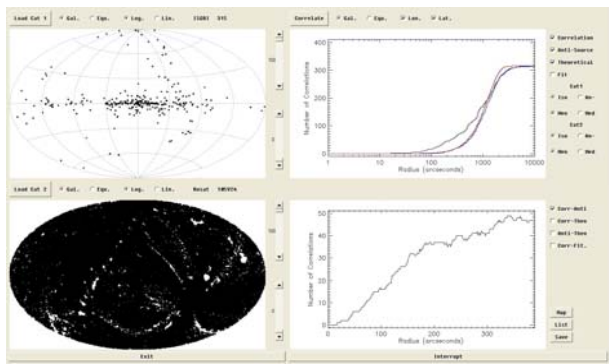

Fig. 2. As figure 1, but using the ROSAT faint source catalogue

### 4.2 Optical: Active Galaxies

At optical wavelengths we use the 11th edition of the Catalogue of Quasars and Active Galactic Nuclei by Véron-Cetty and Véron which contains 48921 quasars, 876 BL Lac objects and 15069 active galaxies (including 11777 Seyfert 1), for a total of 64866 objects [10]. Again a clear correlation is observed with over 50 IBIS/ISGRI objects having a counterpart. The PSLA clearly extends out to about a maximum of 3-4 arcminutes, consistent with the IBIS/ISGRI telescope parameters.

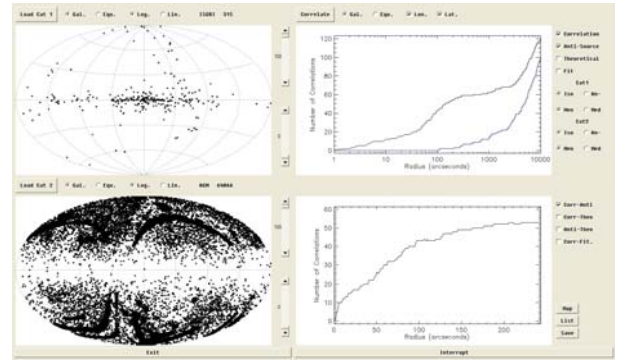

Fig. 3. The ISGRI-AGN correlation

### 4.3 Radio and Infra-red

The combination of high energy emission together with detection in radio and/or infrared is a strong indication of extragalactic nature. For this reason we use the NVSS radio [11] and the extended 2MASS infra-red [12] catalogues. The former contains approximately 2million sources above -40 degrees in declination (the 'hole' below this declination is taken into account in the analysis by filling this area with false sources mirrored from the Northern hemisphere). The latter has about 1.65 million sources, which are primarily galaxies. Although the individual correlations are less striking, when taken together they provide invaluable information with which to help identify the individual sources.

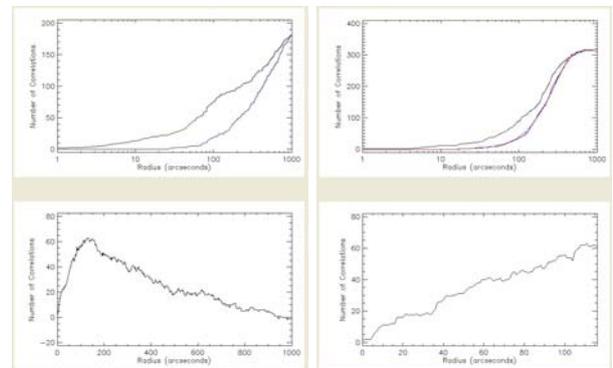

Fig. 4. The correlation function with radio (left) and infra-red (right) sources.

## 5. CONCLUSIONS

We have developed a software tool which allows us to quickly and easily investigate the correlations between various archives in order to help identify sources classified as unknown in the individual IBIS/ISGRI survey catalogues. It has already led to many successful identifications within the first two survey catalogues and will be applied to the 3rd year results as soon as they are available